\documentclass[12pt]{JHEP3}
\usepackage{epsfig}
\epsfclipon
\usepackage{multicol}
\usepackage{epsfig,bm}
\usepackage{amssymb,amsmath}
\usepackage{graphicx}

\newcommand{\be}{\begin{equation}}
\newcommand{\ee}{\end{equation}}
\newcommand{\bea}{\begin{eqnarray}}
\newcommand{\beas}{\begin{eqnarray*}}
\newcommand{\eea}{\end{eqnarray}}
\newcommand{\eeas}{\end{eqnarray*}}
\newcommand{\ba}{\begin{array}}
\newcommand{\ea}{\end{array}}
\def\ls{\mathrel{\lower4pt\vbox{\lineskip=0pt\baselineskip=0pt
           \hbox{$<$}\hbox{$\sim$}}}}
\def\gs{\mathrel{\lower4pt\vbox{\lineskip=0pt\baselineskip=0pt
           \hbox{$>$}\hbox{$\sim$}}}}

\def\smiley{\hbox{\large$\bigcirc$\hspace{-.80em}%
\raise.2ex\hbox{$\cdot\cdot$}\kern-.61em    
\lower.2ex\hbox{\scriptsize$\smile$}}\ }

\newcommand{\roughly}[1]{\mathrel{\raise.3ex\hbox{$#1$\kern-0.85em
\lower1ex\hbox{$\sim$}}}}

\def\be{\begin{equation}}
\def\beq\begin{equation}
\def\ee{\end{equation}}
\def\bea{\begin{eqnarray}}
\def\eea{\end{eqnarray}}

\def\beq{\begin{equation}}
\def\eeq{\end{equation}}
\def\beqa{\begin{eqnarray}}
\def\eeqa{\end{eqnarray}}

\newcommand{\bmat}{\left(\begin{array}}
\newcommand{\emat}{\end{array}\right)}

\title{MSSM flat direction inflation: slow roll, stability, fine
tunning and reheating}

\author{Rouzbeh Allahverdi$^{1}$, Kari Enqvist $^{2}$,
Juan Garcia-Bellido $^{3}$, Asko Jokinen $^{4,5}$, and Anupam
Mazumdar$^{4}$\\
$^{1}$~Perimeter Institute for Theoretical Physics, Waterloo, ON,
N2L 2Y5, Canada \\
$^{2}$~Department of Physical Sciences, University of Helsinki,
and Helsinki Institute of Physics,
P.O. Box 64, FIN-00014 University of Helsinki, Finland \\
$^{3}$~Departamento de F\'\i sica Te\'orica \ C-XI, Universidad
Aut\'onoma de Madrid, Cantoblanco, 28049 Madrid, Spain \\
$^{4}$~NORDITA, Blegdamsvej-17, Copenhagen-2100, Denmark \\
$^{5}$~Laboratoire de Physique Theorique et Astroparticules, UMR 5207 CNRS UM2,
CC 070, Bat 13, Place E. Bataillon, 34095 Montpellier Cedex 5, France.}

\abstract{We consider low scale slow roll inflation driven by the
gauge invariant flat directions {\bf udd} and {\bf LLe} of the
Minimally Supersymmetric Standard Model at the vicinity of a saddle
point of the scalar potential. We study the stability of saddle point
and the slow roll regime by considering radiative and supergravity
corrections. The latter are found to be harmless, but the former
require a modest finetuning of the saddle point condition. We show
that while the inflaton decays almost instantly, full thermalization
occurs late, typically at a temperature $T\approx 10^{7}$~GeV, so that
there is no gravitino problem. We also compute the renormalization
group running of the inflaton mass and relate it to slepton masses
that may be within the reach of LHC and could be precisely determined
in a future Linear Collider experiment.}

\begin{document}


\section{Introduction }

Recently we have constructed a model of inflation~\cite{AEGM} based
on the {\bf udd} and {\bf LLe} flat directions of Minimally
Supersymmetric Standard Model (MSSM; for a review of MSSM flat
directions, see~\cite{KARI-REV}). In this model the inflaton is a
gauge invariant combination of either squark or slepton fields.  For
a choice of the soft SUSY breaking parameters $A$ and the inflaton
mass $m_\phi$, the potential along the flat {\bf udd} and {\bf LLe}
directions is such that there is a period of slow roll inflation of
sufficient duration to provide the observed spectrum of CMB
perturbations. In the inflationary part of the MSSM potential the
second derivative is vanishing and the slow roll phase is driven by
the third derivative of the potential~\footnote{In a recent similar
model with small Dirac neutrino masses, the observed microwave
background anisotropy and the tilted power spectrum are related to
the neutrino masses~\cite{AKM}. The model relies solely on
renormalizable couplings.}.

MSSM inflation occurs at a very low scale with $H_{inf}\sim 1-10$~GeV
and with field values much below the Planck scale. Hence it stands in
strong contrast to the conventional inflation models which are based
on ad hoc gauge singlet fields and often employ field values close to
Planck scale (for a review, see ~\cite{LYTH}). In such models the
inflaton couplings to SM physics are unknown.  As a consequence, much
of the post-inflationary evolution, such as reheating, thermalization,
generation of baryon asymmetry and cold dark matter, which all depend
very crucially on how the inflaton couples to the (MS)SM
sector~\cite{AVERDI1,AVERDI2,AVERDI3,AVERDI4}, is not calculable from
first principles. The great virtue of MSSM inflation based on flat
directions is that the inflaton couplings to Standard Model particles
are known and, at least in principle, measurable in laboratory
experiments such as LHC or a future Linear Collider.

However, as in almost all inflationary models, a fine tuning of the
initial condition is needed to place the flat direction field $\phi$
to the immediate vicinity of the saddle point $\phi_0$ at the onset of
inflation. In addition, there is the question of the stability of the
saddle point solution and of the existence of a slow roll regime.
These are issues that we wish to address in detail in the present
paper. Both supergravity and radiative corrections to the flat
direction inflaton potential must be considered.  Hence we need to
write down and solve the renormalization group (RG) equations for the
MSSM flat directions of interest. RG equations are also needed to
scale the model parameters, such as the inflaton mass, down to TeV
scale; since the inflaton mass is related either to squark or slepton
masses, it could be measured by LHC or a future Linear Collider.

Because the inflaton couplings to ordinary matter are known, inflaton
decay and thermalization are processes that can be computed in an
unambiguous way. Unlike in many models with a singlet inflaton, in MSSM
inflation the potential relevant for decay and thermalization cannot
be adjusted independently of the slow roll part of the potential.

This paper is organized as follows. In Sect. 2 we present the model of
MSSM inflation and its predictions. In Sect. 3 we study the flat
direction potential without an exact saddle point. We find generic
constraints for the existence of a slow roll solution and show that in
the slow roll regime there is always tunneling from a false minimum.
In Sect. 4 we solve the one-loop RG equations for the {\bf LLe} flat
direction and find the one-loop corrected saddle point. We quantify
the amount of fine tuning required for the slow roll solution to
exist, and relate through RG running the {\bf LLe} inflaton mass with
observables such as the slepton masses at the LHC energy scale.  We
also show that supergravity corrections to the potential can be
neglected.  In Sect. 5 we discuss the decay of the flat direction, the
reheating and thermalization of the Universe, and show that the reheat
temperature is low enough for the model to avoid the gravitino
problem.  Sect. 6 contains our conclusions and some discussion about
future prospects.


\section{The Model}\label{model}

Let us recapitulate the main features of MSSM flat direction
inflation~\cite{AEGM}. As is well known, in the limit of unbroken
SUSY the flat directions have exactly vanishing potential. This
situation changes if we take into account soft SUSY breaking and
non-renormalizable superpotential terms~\footnote{Our framework is
MSSM together with gravity, so consistency dictates that all
non-renormalizable terms allowed by gauge symmetry and supersymmetry
should be included below the cut-off scale, which we take to be the
Planck scale. Some interseting issues on A-term inflation were also
discussed in Ref.~\cite{LYTH0}.} of the type~\cite{KARI-REV}

\beq \label{supot}
W_{non} = \sum_{n>3}{\lambda_n \over n}{\Phi^n \over M^{n-3}}\,,
\eeq
where $\Phi$ is a
superfield which contains the
flat direction.  Within MSSM all the flat directions are lifted by
non-renormalizable operators with $4\le n\le 9$~\cite{GKM}, where $n$
depends on the flat direction. We expect that quantum gravity effects yield
$M=M_{\rm P}=2.4\times
10^{18}$~GeV and $\lambda_n\sim {\cal
O}(1)$~\cite{DRT}~\footnote{Note however that our results will be valid
for any values of $\lambda_n$, because rescaling $\lambda_n$ simply shifts
the VEV of the flat direction.}.


\begin{figure}
\vspace*{-0.0cm}
\begin{center}
\epsfig{figure=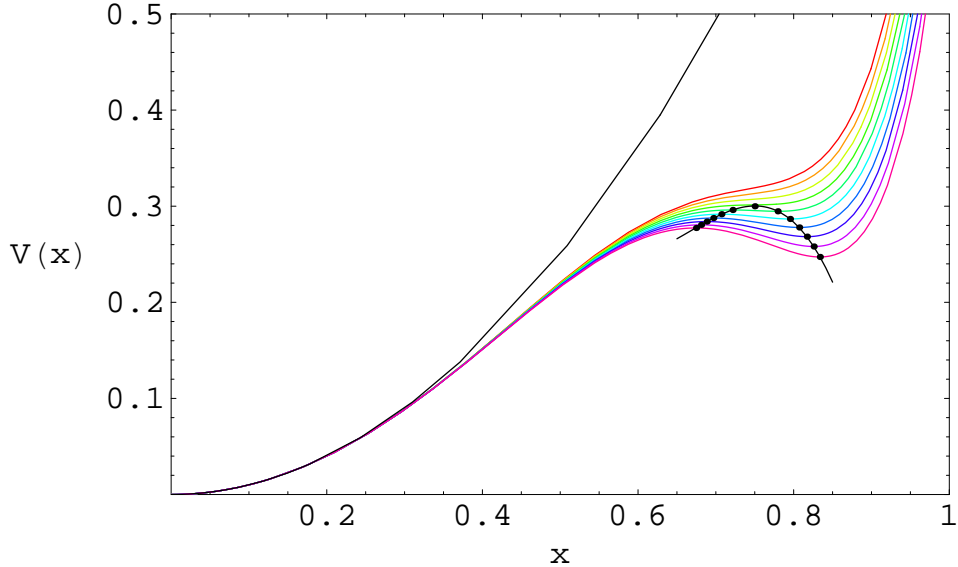,width=.84\textwidth,clip=}
\vspace*{-0.0cm}
\end{center}
\caption{ The colored curves depict the full potential, where
$V(x)\equiv V(\phi)/(0.5~m_{\phi}^2 M_{\rm P}^2(m_{\phi}/M_{\rm
P})^{1/2})$, and $ x\equiv (\lambda_n M_{\rm P}/m_{\phi})^{1/4}
(\phi/M_{\rm P})$. The black curve is the potential arising from the
soft SUSY breaking mass term. The black dots on the colored potentials
illustrate the gradual transition from minimum to the saddle point and
to the maximum.}
\label{fig0-pot}
\end{figure}


Let us focus on the lowest order superpotential term in
Eq.~(\ref{supot}) which lifts the flat direction. Soft SUSY breaking
induces a mass term for $\phi$ and an $A$-term so that the scalar
potential along the flat direction reads
\beq \label{scpot}
V = {1\over2} m^2_\phi\,\phi^2 + A\cos(n \theta  + \theta_A)
{\lambda_{n}\phi^n \over n\,M^{n-3}_{\rm P}} + \lambda^2_n
{{\phi}^{2(n-1)} \over M^{2(n-3)}_{\rm P}}\,,
\eeq
Here $\phi$ and $\theta$ denote respectively the radial and the
angular coordinates of the complex scalar field
$\Phi=\phi\,\exp[i\theta]$, while $\theta_A$ is the phase of the
$A$-term (thus $A$ is a positive quantity with dimension of mass).
Note that the first and third terms in Eq.~(\ref{scpot}) are positive
definite, while the $A$-term leads to a negative contribution along
the directions whenever $\cos(n \theta + \theta_A) < 0$.~\footnote{The
importance of the A-term has also been highlighted in a successful
MSSM curvaton model~\cite{AEJM}.}

In principle, in the $A$-term all the superpotential terms of a given
dimension $n$ may enter with a different coefficient $A_n$; whether
they are related or not depends on the details of the SUSY breaking
mechanism.


\subsection{The Saddle Point}

The maximum impact from the $A$-term is obtained when $\cos(n \theta +
\theta_A) = -1$ (which occurs for $n$ values of $\theta$).  Along
these directions $V$ has a secondary minimum at $\phi = \phi_0 \sim
\left(m_\phi M^{n-3}_{\rm P}\right)^{1/n-2} \ll M_{\rm P}$ (the global
minimum is at $\phi=0$), provided that
\beq \label{extrem}
A^2 \geq 8 (n-1) m^2_{\phi}\,.
\eeq
At this minimum the curvature of the potential is positive both along
the radial and angular directions\footnote{If the $A$ is too large,
the secondary minimum will be deeper than the one in the origin, and
hence becomes the true minimum. However, this is {\it
phenomenologically} unacceptable as such a minimum will break charge
and/or color~\cite{DRT}. } with $V \sim m_\phi^2\phi_0^2 \sim
m_{\phi}^2\left(m_{\phi} M^{n-3}_{\rm P}\right)^{2/(n-2)}$.

As discussed in ~\cite{AEGM}, if the local minimum is too steep, the
field will become trapped there with an ensuing inflation that has no
graceful exit like in the old inflation scenario~\cite{GUTH}.  On the
other hand in an opposite limit, with a point of inflection, a single
flat direction cannot support inflation~\cite{JOKINEN}.

However, in the gravity mediated SUSY breaking case, the $A$-term and
the soft SUSY breaking mass terms are expected to be the same order of
magnitude as the gravitino mass, i.e.
\begin{eqnarray}
\label{condi-0}
m_{\phi}\sim A \sim m_{3/2}\sim {\cal O}(1)~{\rm TeV}\,.
\end{eqnarray}
Therefore, as pointed out in~\cite{AEGM}, in the gravity mediated SUSY
breaking it is possible that the potential barrier actually disappears
and the inequality in Eq.~(\ref{extrem}) is saturated so that $A$ and
$m_\phi$ are related by
\beq
\label{cond}
A^2 = 8 (n-1) m^2_\phi\,.
\eeq
This represents a fine tuning and will be discussed at length in the
next Sections. However, let us now assume for the sake of argument
that Eq.~(\ref{cond}) holds.  Then both the first and second
derivatives of $V$ vanish at $\phi_0$, ~i.e. $V^{\prime}(\phi_0)=0,~
V^{\prime\prime}(\phi_0)=0$.  As the result, if initially $\phi\sim
\phi_0$, a slow roll phase of inflation is driven by the third
derivative of the potential.

Note that this behavior does not seem possible for other SUSY
breaking scenarios such as the gauge mediated breaking~\cite{GMSB} or
split SUSY~\cite{SPLIT}. In split SUSY the $A$-term is protected by an
$R$-symmetry, which also keeps the gauginos light while the sfermions
are quite heavy~\cite{SPLIT}~\footnote{In the gauge mediated case
there is an inherent mismatch between $A$ and $m_{\phi}$, except at
very large field values where Eq.~(\ref{condi-0}) can be satisfied.
However there exists an unique possibility of a saddle point inflation
which we will discuss separately~\cite{AJKM}.}.


\subsection{Slow roll}

The potential near the saddle point Eq. (\ref{cond}) is very flat
along the {\it real direction} but not along the {\it imaginary
direction}. Along the imaginary direction the curvature is determined
by $m_{\phi}$.  Around $\phi_0$ the field lies in a plateau with a
potential energy
\beq \label{potential}
V(\phi_0) = {(n-2)^2\over2n(n-1)}\,m^2_\phi \phi_0^2
\eeq
with
\beq \label{phi0}
\phi_0 = \left({m_\phi M^{n-3}_{\rm P}\over
\lambda_n\sqrt{2n-2}}\right)^{1/(n-2)}\,.
\eeq
This results in Hubble expansion rate during inflation which is given by
\beq \label{hubble}
H_{\rm inf} = {(n-2) \over \sqrt{6 n (n-1)}} {m_{\phi} \phi_0 \over M_{\rm P}}.
\eeq
When $\phi$ is very close to $\phi_0$, the first derivative is
extremely small. The field is effectively in a de Sitter background,
and we are in self-reproduction (or eternal inflation) regime where the
two point correlation function for the flat direction fluctuation
grows with time. But eventually classical friction wins and slow roll
begins at $\phi \approx \phi_{\rm self}$~\cite{AEGM}
\beq \label{self}
(\phi_0-\phi_{\rm self}) \simeq \Big({m_\phi \phi_0^2 \over M_{\rm
P}^3}\Big)^{1/2} \phi_0.
\eeq
The slow roll potential in this case reads
\begin{eqnarray} \label{potential2}
&&V(\phi) = V(\phi_0) + {1\over3!} V'''(\phi_0)
(\phi-\phi_0)^3 +\cdot\cdot\cdot \,, \nonumber \\
\label{3rder}
&&V^{\prime \prime \prime}({\phi_0}) = 2(n-2)^2
{m^2_\phi \over \phi_0}\,.
\end{eqnarray}
We can now solve the equation of motion for the $\phi$ field in the
slow-roll approximation,
\beq \label{slow}
3H\dot\phi=-\frac{1}{2}V'''(\phi_0)(\phi-\phi_0)^2\,,
\eeq
assuming initial conditions such that the flat direction starts in the
vicinity of $\phi_0$ with $\dot\phi\approx 0$.  Inflation ends when
the slow roll parameter, $\epsilon\equiv (M_{\rm P}^2/2)
(V^{\prime}/V)^2$ becomes of ${\cal O}(1)$. This occurs at
\beq \label{end}
(\phi_0-\phi_{\rm end}) \sim {\phi^3_0 \over 2n(n-1)M^2_{\rm P}}\,.
\eeq
which happens to be also the place when the other slow roll
paremeter $\eta \equiv M^2_{\rm P}(V^{\prime \prime}/V)$ becomes
of ${\cal O}(1)$.

The number of e-foldings during the slow roll from $\phi$ to
$\phi_{\rm end}$ is given by
\beq \label{efold}
{\cal N}_e(\phi) = \int_{\phi}^{\phi_{\rm end}} {H_{\rm inf} d\phi
\over \dot\phi} \simeq {\phi^3_0 \over 2n(n-1)M^2_{\rm P}(\phi_0 - \phi)}\,,
\eeq
where we have used $V'(\phi) \sim (\phi - \phi_0)^2 V'''(\phi_0)$
(this is justified since $V'(\phi_0) \sim 0, V''(\phi_0)\sim 0$), and
Eq.~(\ref{slow}). The total number of e-foldings in the slow roll
regime is then found from Eq.(\ref{self})
\beq \label{tot}
{\cal N}_{\rm tot} \simeq {1 \over 2n(n-1)} \big({\phi^2_0 \over m_{\phi}
M_{\rm P}}\Big)^{1/2}.
\eeq
The observationally relevant perturbations are generated when $\phi
\approx \phi_{\rm COBE}$. The number of e-foldings between $\phi_{\rm
COBE}$ and $\phi_{\rm end}$, denoted by ${\cal N}_{\rm COBE}$ follows
from Eq.~(\ref{efold})
\beq \label{cobe}
{\cal N}_{\rm COBE} \simeq {\phi^3_0 \over 2n(n-1)M^2_{\rm P}(\phi_0 -
\phi_{\rm COBE})}.
\eeq
The amplitude of perturbations thus produced is given by
\beq \label{ampl}
\delta_{H} \equiv \frac{1}{5\pi}\frac{H^2_{\rm inf}}{\dot\phi} \simeq
\frac{1}{5\pi} \sqrt{\frac{2}{3}n(n-1)}(n-2) ~ \Big({m_\phi M_{\rm P} \over
\phi_0^2}\Big) ~ {\cal N}_{\rm COBE}^2,
\eeq
where we have used Eqs.(\ref{hubble}), (\ref{potential2}),
(\ref{cobe}). Again after using these equations, the spectral tilt of
the power spectrum and its running are found to be
\begin{eqnarray}
\label{tilt}
&&n_s = 1 + 2\eta - 6\epsilon \ \simeq \ 1 -
{4\over {\cal N}_{\rm COBE}} \,, \\ \label{running}
&&{d\,n_s\over d\ln k} = - {4\over {\cal N}_{\rm COBE}^2}. \,
\end{eqnarray}
%

\subsection{Properties and predictions}

As discussed in~\cite{AEGM}, among the about 300 flat directions there
are two that can lead to a successful inflation along the lines
discussed above.

One is {\bf udd} which, up to an overall phase factor, is parameterized by
\beq
\label{example}
u^{\alpha}_i=\frac1{\sqrt{3}}\phi\,,~
d^{\beta}_j=\frac1{\sqrt{3}}\phi\,,~
d^{\gamma}_k=\frac{1}{\sqrt{3}}\phi\,.
\eeq
Here $1 \leq \alpha,\beta,\gamma \leq 3$ are color indices, and $1
\leq i,j,k \leq 3$ denote the quark families. The flatness constraints
require that $\alpha \neq \beta \neq \gamma$ and $j \neq k$.

The other direction is {\bf LLe}, parameterized by (again up to an overall
phase factor)
\beq
L^a_i=\frac1{\sqrt{3}}\left(\begin{array}{l}0\\ \phi\end{array}\right)\,,~
L^b_j=\frac1{\sqrt{3}}\left(\begin{array}{l}\phi\\ 0\end{array}\right)\,,~
e_k=\frac{1}{\sqrt{3}}\phi\,,
\eeq
where $1 \leq a,b \leq 2$ are the weak isospin indices and $1 \leq
i,j,k \leq 3$ denote the lepton families. The flatness constraints
require that $a \neq b$ and $i \neq j \neq k$.  Both these flat
directions are lifted by $n=6$ non-renormalizable operators,
\begin{eqnarray}
W_6\supset\frac{1}{M_{\rm P}^3}(LLe)(LLe)\,,\hspace{1cm}
W_6\supset \frac{1}{M_{\rm P}^3}(udd)(udd)\,.
\end{eqnarray}
The reason for choosing either of these two flat
directions\footnote{Since {\bf LLe} are {\bf udd} are independently
$D$- and $F$-flat, inflation could take place along any of them but
also, at least in principle, simultaneously. The dynamics of multiple
flat directions are however quite involved~\cite{EJM}.} is twofold:
(i) a non-trivial $A$-term arises, at the lowest order, only at $n=6$;
and (ii) we wish to obtain the correct COBE normalization of the CMB
spectrum.

Those MSSM flat directions which are lifted by operators with
dimension $n=7,9$ are such that the superpotential term contains at
least two monomials, i.e. is of the type
\begin{eqnarray}\label{doesnotcontri}
W \sim \frac{1}{M_{\rm P}^{n-3}}\Psi\Phi^{n-1}\,.
\end{eqnarray}
If $\phi$ represents the flat direction, then its VEV induces a large
effective mass term for $\psi$, through Yukawa couplings, so that
$\langle \psi \rangle =0$. Hence Eq. (\ref{doesnotcontri}) does not
contribute to the $A$-term.

More importantly, as we will see, all other flat directions except
those lifted by $n=6$ fail to yield the right amplitude for the
density perturbations. Indeed, as can be seen in Eq.~(\ref{phi0}), the
value of $\phi_0$, and hence also the energy density, depend on $n$.

According to the arguments presented above, successful MSSM flat direction
inflation has the following model parameters:
\beq
m_{\phi}\sim 1-10~{\rm TeV}\,,~~n=6\,,~~A=\sqrt{40}m_{\phi}\,,
~~\lambda\sim {\cal O}(1)\,.
\label{VALVS}
\eeq
Here we assume that $\lambda$ (we drop the subscript "6") is of order
one, which is the most natural assumption when $M=M_{\rm P}$.

The Hubble expansion rate during inflation and the VEV of the saddle
point are~\footnote{We note that $H_{\rm inf}$ and $\phi_0$ depend
very mildly on $\lambda$ as they are both $\propto \lambda^{-1/4}$.}
\beq \label{values}
H_{\rm inf}\sim 1-10~{\rm GeV}\,,~~~\phi_0 \sim (1-3) \times
10^{14}~{\rm GeV}\,.
\eeq
Note that both the scales are sub-Planckian. The total energy density
stored in the inflaton potential is $V_0 \sim 10^{36}-10^{38}~{\rm
GeV}^4$. The fact that $\phi_0$ is sub-Planckian guarantees that the
inflationary potential is free from the uncertainties about physics at
super-Planckian VEVs. The total number of e-foldings during the slow
roll evolution is large enough to dilute any dangerous relic away, see
Eq.~(\ref{tot}):
\beq \label{totalefold}
{\cal N}_{\rm tot} \sim 10^3  \,,
\eeq
Domains which are initially closer than $\phi_{\rm self}$ to $\phi_0$,
see Eq.~(\ref{self}), can enter self-reproduction in eternal
inflation, with no observable consequences.

At such low scales as in MSSM inflation the number of e-foldings,
${\cal N}_{\rm COBE}$, required for the observationally relevant
perturbations, is much less than $60$~\cite{MULTI}.  If the inflaton
decays immediately after the end of inflation, we obtain ${\cal
N}_{\rm COBE} \sim 50$. Despite the low scale, the flat direction can
generate adequate density perturbations as required to explain the
COBE normalization. This is due to the extreme flatness of the
potential (recall that $V'=0$), which causes the velocity of the
rolling flat direction to be extremely small. From Eq.~(\ref{ampl}) we
find an amplitude of
\beq
\label{amp}
\delta_{H} \simeq 1.91 \times 10^{-5}\,.
\eeq

There is a constraint on the mass of the flat direction from the
amplitude of the CMB anisotropy:
\begin{equation}
\label{mbound} m_{\phi} \simeq (100 ~ {\rm GeV}) \times \lambda^{-1}
\, \left( \frac{{\cal N}_{\rm COBE}}{50} \right)^{-4}\,.
\end{equation}
We get a lower limit on the mass parameter when $\lambda\leq 1$.
For smaller values of $\lambda\ll 1$, the mass of the flat
direction must be larger.  Note that the above bound on the inflaton
mass arises at high scales, i.e. $\phi=\phi_0$. However, through
renormalization group flow, it is connected to the low scale mass, as
will be discussed in Sect. 4.

The spectral tilt of the power spectrum is not negligible because,
although the first slow roll parameter is $\epsilon\sim1/{\cal
N}_{\rm COBE}^4\ll 1$, the other slow roll parameter is given by
$\eta = -2/{\cal N}_{\rm COBE}$ and thus, see
Eq.~(\ref{tilt})\footnote{Obtaining $n_s > 0.92$ (or $n_s < 0.92$,
which is however outside the $2 \sigma$ allowed region) requires
deviation from the saddle point condition in Eq.~(\ref{cond}), see
Section 3. For a more detailed discussion on the spectral tilt, see
also Refs.~\cite{LYTH1},\cite{Maz-Roz}.}
\begin{eqnarray}
\label{spect}
&&n_s
\sim 0.92\,,\\
&&{d\,n_s\over d\ln k}
\sim - 0.002\,,
\end{eqnarray}
where we have taken ${\cal N}_{\rm COBE} \sim 50$ (which is the
maximum value allowed for the scale of inflation in our model). In the
absence of tensor modes, this agrees with the current WMAP 3-years'
data within $2\sigma$~\cite{WMAP3}. Note that MSSM inflation does not
produce any large stochastic gravitational wave background during
inflation. Gravity waves depend on the Hubble expansion rate, and in
our case the energy density stored in MSSM inflation is very small.


\section{Sensitivity of the saddle point inflation}

In previous Sections and in Ref.~\cite{AEGM} the dynamics of the
flat direction inflaton was discussed assuming the saddle point
condition Eq.~(\ref{cond}) is satisfied exactly.  The question then
is, how large a deviation can be allowed for before slow roll
inflation will be spoiled. There are obviously two distinct
possibilities: either $A>\sqrt{8(n-1)}m_{\phi}$ or
$A<\sqrt{8(n-1)}m_{\phi}$. (Although we always take $n=6$ in the
present paper, we keep $n$ here for generality of the formalism.) In
the former case there is a barrier which separates the global
minimum $\phi=0$ and the false minimum at $\phi\simeq\phi_0$. The
eventual inflationary trajectory starts near the top of the barrier.
The field can either start at the top, or jump to its vicinity from
the false minimum via Coleman-de Luccia tunneling~\cite{COLEMAN}. As
we will see, if the barrier is too high, there will be no inflation
near the top. In the latter case there is no minimum but the
potential may be too steep for slow roll inflation. Therefore we
need to analyze the two cases separately. However, the steepness of
the potential is a problem which is common to both cases and is
addressed at the end of this Section.

To facilitate the discussion, let us define
\begin{eqnarray}
\delta\equiv {A^2\over8(n-1)m_\phi^2}
\equiv 1 \pm \left({n-2\over2}\right)^2\,\alpha^2\,.
\end{eqnarray}
Here we will assume that $\alpha \ll 1$. Before beginning the
calculations, we would like to point that the main results of this
section are summarized in Fig.~(\ref{tiltbeta}) and
Eq.~(\ref{fine}). These yield no constraint on the spectral tilt as
any value consistent with sufficient slow roll inflation (i.e. a
number $\geq {\cal N}_{\rm COBE}$ of e-foldings) is allowed.


\subsection{The potential for $\delta>1$ }

In this case there are two extrema, a maximum $(-)$ and a minimum
$(+)$,
\begin{equation}\label{minmax}
\phi_\pm = \phi_0 \,\left[\sqrt\delta \pm
\sqrt{\delta -1}\right]^{1\over n-2} =
\phi_0\left(1\pm{\alpha\over2}+{\cal O}(\alpha^2)\right)\,,
\end{equation}
and a point of inflection $\phi_i$
\beq \label{inf}
\phi_i = \phi_0 \Big(1 + {n-2 \over 32} \alpha^2 + {\cal O}(\alpha^4)\Big).
\eeq
We can then express the potential and its derivatives at the extrema
$\phi_{-}$ and $\phi_{+}$ as functions of $\alpha$,
\begin{eqnarray}
&&V(\phi_\pm) = V(\phi_0)\left(1-{n-1\over2}\alpha^2
\pm {n(n-1)\over6}\alpha^3\right)\,,\label{potmin}\\
&&V''(\phi_\pm) = \pm \alpha (n-2)^2 m_\phi^2 \,,\label{masses}\\
&&V'''(\phi_\pm) = V'''(\phi_0)\left(1 \pm {3\over2}(n-2)\alpha\right) \,.
\end{eqnarray}
Note that $\alpha\to0$ when $\delta\to1$, so we can expand the
potential around the maximum, and include the small $\alpha$
correction due to deviations from the saddle point as
\begin{equation}\label{Vsaddle}
V(\phi) = (n-2)^2 m_\phi^2 \phi_0^2\left[{1\over2n(n-1)} -
{\alpha\over2}\left({\phi\over\phi_0}-1+{\alpha\over2}\right)^2 +
{1\over3}\left({\phi\over\phi_0}-1+{\alpha\over2}\right)^3 \right]\,.
\end{equation}
The maximum is now at $\phi_{\rm max}=\phi_0(1-\alpha/2)$, and the
minimum is at $\phi_{\rm min}=\phi_0(1+\alpha/2)$, with masses
(curvature of the potential) given by $m^2_\pm=\pm(n-2)^2m_\phi^2\,
\alpha$, which coincides with Eq.~(\ref{masses}), while there is now a
point of inflection at $\phi_0$. Note that the difference in potential
height between the maximum and the minimum is
\begin{equation}\label{DV}
{\Delta V\over V_0} = {V_{\rm max} - V_{\rm min}\over V(\phi_0)} =
{n(n-1)\over3}\alpha^3\,.
\end{equation}
In the limit of $\alpha\to0$ we recover the saddle point. We will work
in the limit when $\alpha \ll 1$.


Let us now define a few variables, $x=\phi/\phi_0$,
$\,\tau=(n-2)m_\phi t$ \ and \ $h=H_{\rm inf}/(n-2)m_\phi$. Then the
equation of motion for the scalar field down the potential can be
written as
\begin{equation}\label{eqmotion}
x'' + 3h x' + V'(x) = x'' + 3h x' + \Big(x-1+{\alpha\over2}\Big)^2 -
\alpha\Big(x-1+{\alpha\over2}\Big) = 0\,,
\end{equation}
where we have used (\ref{Vsaddle}).

The eventual inflationary trajectory will start in the vicinity of
the maximum $x=1-\alpha/2$ and will roll down the hill towards
$x=0$. The field can either start near the maximum, or tunnel to its
vicinity out of the false vacuum. Tunneling takes place in the
presence of a non-zero vacuum energy, $H_{\rm inf}\neq0$, and is
known as Coleman-de Luccia tunneling~\cite{COLEMAN}.

In order to find the interpolating solution between the false and the
true vacuum one solves the {\em Euclidean} equation of motion,
\begin{equation}\label{euclideaneqmotion}
x'' + 3h x' - V'(x) = x'' + 3h x' - \Big(x-1+{\alpha\over2}\Big)^2 +
\alpha\Big(x-1+{\alpha\over2}\Big) = 0\,,
\end{equation}
whose exact solution is
\begin{equation}\label{solutionx}
x(\tau) = 1 - {\alpha\over2} \tanh{\alpha\tau\over6h}\,.
\end{equation}
This solution starts at $x(-\infty)=1+\alpha/2=\phi_+/\phi_0$ and ends
at $x(\infty)=1-\alpha/2=\phi_-/\phi_0$.  The ``tunneling'' from
$\phi_{\rm min}$ to $\phi_{\rm max}$ can actually be understood as
diffusion due to de Sitter fluctuations.  It is valid so long as
$V^{\prime \prime}(\phi_{\rm min}) \ll H^2_{\rm inf}$. This requires
that, see Eq.~(\ref{masses}),
\begin{equation}\label{tunneling}
\alpha \ll {1\over6n(n-1)} {\phi_0^2\over M_{\rm P}^2}\,.
\end{equation}
This also insures that $|\eta| \ll 1$ at
the maximum, and hence inflation can take place after
tunneling. Otherwise there will be no inflation, neither in
self-reproduction nor in slow roll regime.

\begin{figure}
\vspace*{-0.0cm}
\begin{center}
\epsfig{figure=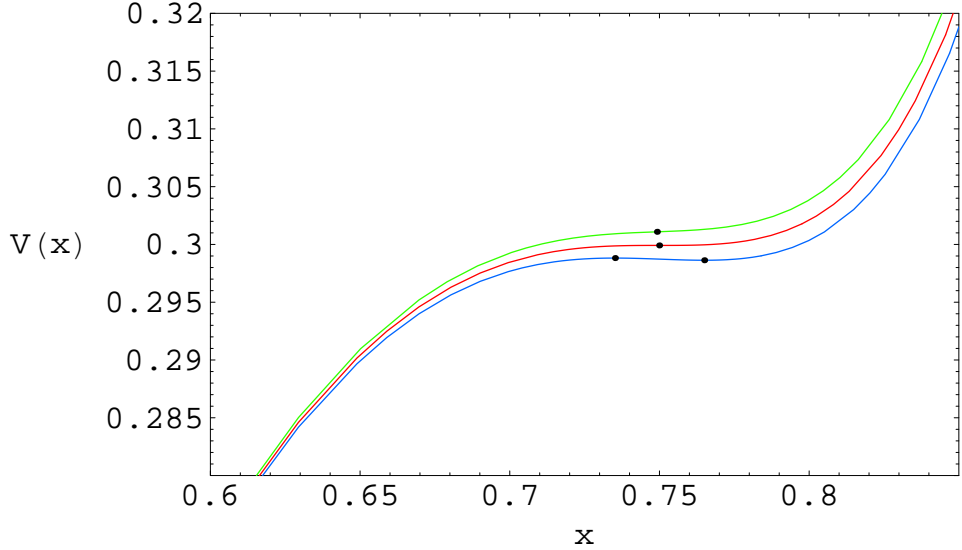,width=.84\textwidth,clip=}
\vspace*{-0.0cm}
\end{center}
\caption{ The red curve depicts the potential, $V(x)\equiv
V(\phi)/(0.5m_{\phi}^2M_{\rm P}^2(m_{\phi}/M_{\rm P})^{1/2})$, where
$x\equiv (\lambda_n M_{\rm P}/m_{\phi})^{1/4}(\phi/M_{\rm P})$, for
the saddle point (shown by the black dot) when $\delta=1$. The blue
curve illustrates the potential when $\delta =1+\sqrt{40}/1000$, where
the two black dots, one on right shows the minimum value $\phi_-$ and
on the left shows the maximum value, $\phi_{+}$.  The green curve
portrays the potential for the opposite case when $\delta
=1-\sqrt{40}/1000$. The black dot is the point of inflection.}
\label{fig-pot}
\end{figure}


Let us now discuss the effect of the tunneling solutions on the tilt
of the CMB spectrum. Again there is self-reproduction close to the
maximum as long as the curvature there is smaller than the rate of
expansion squared, i.e.  $\alpha \ll 1/180(\phi_0/M_{\rm P})^2$.  The
slow-roll regime starts at $\phi \approx \phi_{\rm self}$ when
\begin{equation}\label{sr}
|(\phi_{\rm self}-\phi_{-})(\phi_{\rm self}-\phi_{+})| \leq
{m_\phi\phi_0^4\over M_{\rm P}^3}\,,
\end{equation}
Note that $\vert(\phi_{\rm self}-\phi_{-})\vert \simeq \
\vert(\phi_{+}-\phi_{-})\vert = \alpha \phi_0$, see Eq.~(\ref{potmin}), and
$\vert(\phi_{\rm self} - \phi_{+})\vert \simeq \vert(\phi_{\rm self} -
\phi_{0})\vert$. We therefore find
\beq \label{self2}
(\phi_{0} - \phi_{\rm self}) \simeq {m_{\phi} \phi^4_0 \over
\alpha M^3_{\rm P}}.
\eeq

Now we integrate Eq.~(\ref{eqmotion}) in the slow-roll approximation,
using a new variable $u=x-x_{-}$, for which the equation of motion
becomes $3hu' = - u(u-\alpha)$.  The exact solution is, in terms of the
number of $e$-folds, ${\cal N} = h\tau$,
\begin{equation}\label{Ntau}
x({\cal N}) = 1 - {\alpha\over 2}\,\coth(\beta{\cal N})
\end{equation}
where we have defined
\begin{equation}\label{beta}
\beta = {\alpha\over6h^2} = n(n-1)\,\alpha\,{M_{\rm P}^2\over\phi_0^2}\,.
\end{equation}
Note that in the limit $\alpha\to0$, we recover the usual expression
Eq.~(\ref{efold}). From Eqs.~(\ref{end}),~(\ref{self2}),~(\ref{Ntau}), it
turns out that the number of e-folds from $\phi_{\rm self}$ to the end
of inflation at $\phi_{\rm end}$ is again of order $10^3$.

The required number of e-folds for the relevant perturbations (${\cal
N}_{\rm COBE}\sim 50$) determines the value of $\phi_{\rm COBE}$,
\begin{equation}\label{N40}
|(\phi_{\rm COBE}-\phi_{-})(\phi_{\rm COBE}-\phi_{+})|^{1/2} =
{\alpha\over2\sinh\beta{\cal N}} \simeq
{3h^2\phi_0\over{\cal N}_{\rm COBE}} =
{\phi_0^3\over 2n(n-1)M_{\rm P}^2\,{\cal N}_{\rm COBE}}\,.
\end{equation}

\begin{figure}
\begin{center}
\epsfig{figure=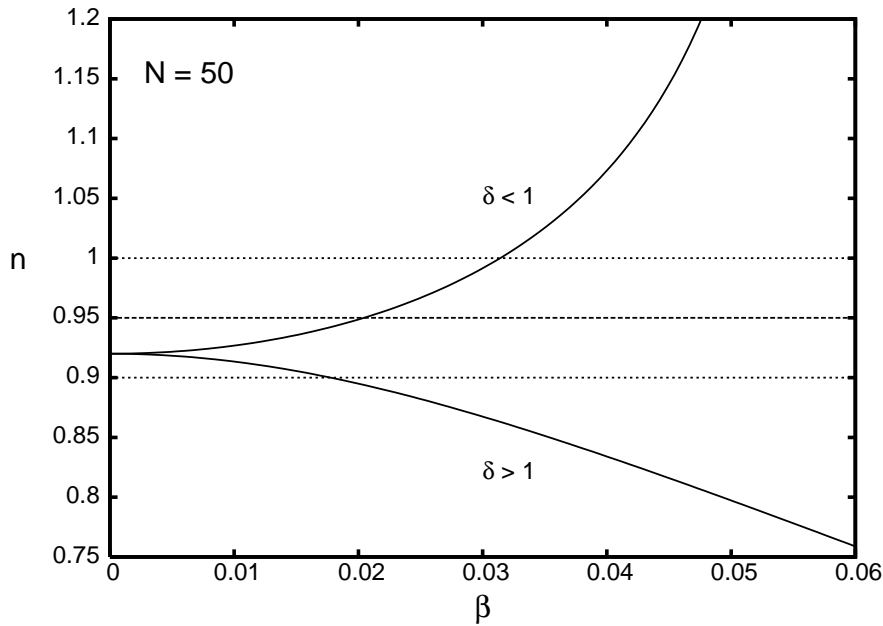,angle=-90,width=.84\textwidth,clip=}
\end{center}
\caption{The tilt of the MSSM model as a function of the saddle point
deviation parameter $\beta$ (3.16). We have plotted both cases
$\delta < 1$ and $\delta > 1$. Note the allowed $2\sigma$ range coming
from a combination of CMB and LSS observations.} \label{tiltbeta}
\end{figure}

On the other hand, the amplitude of fluctuations is given by
\begin{equation}\label{deltaH}
\delta_H \simeq {3\over5\pi}\,{H_{\rm inf}^3\over V'(\phi)} =
{(n-2)\sqrt{n}\,(2n-2)^{n\over2n-4}\over5\pi\sqrt3}\,
\lambda_n^{2\over n-2}\,\left({m_\phi\over M_{\rm P}}\right)^{n-4\over n-2}\,
{\sinh^2\beta{\cal N}\over\beta^2}
\end{equation}
which for $n=6,4,3$ becomes
\begin{eqnarray}\nonumber
&&\delta_H(n=6) \simeq 2\,\left({\lambda_n m_\phi\over M_{\rm P}}
\right)^{1/2}\,
{\sinh^2\beta{\cal N}\over\beta^2}\,,\\[2mm]
&&\delta_H(n=4) \simeq \lambda_n\,
{\sinh^2\beta{\cal N}\over\beta^2}\,,\\[2mm]\nonumber
&&\delta_H(n=3) \simeq 0.5\,\lambda_n^2 {M_{\rm P}\over m_\phi}\,
{\sinh^2\beta{\cal N}\over\beta^2}\,,
\end{eqnarray}
while the spectral tilt and its running are universal,
\begin{eqnarray}\label{tiltN}
&&n_s = 1 - {d\ln\delta_H^2\over d{\cal N}} =
1 - 4\beta\,\coth\beta{\cal N}\,,\\[2mm]\label{runningN}
&&{d\,n_s\over d\,\ln k} = - {d\,n_s\over d{\cal N}} = - {4\beta^2
\over\sinh^2\beta{\cal N}}\,,
\end{eqnarray}
which reduce to the usual expressions in the limit $\beta\to0$, see
Eqs.~(\ref{tilt}),~(\ref{running}). We show in Fig.~\ref{tiltbeta}
the variation of the tilt with $\beta$ for a model with ${\cal
N}_{\rm COBE} = 50$. Note that the range of allowed values of
$\beta$ is constrained by the condition to have inflation near the
maximum, i.e. that $\vert \eta \vert < 1$ at $\phi_{-}$. This gives
$\beta < 0.06$, for $n=6$, see Eqs.~(\ref{tunneling},\ref{beta}).
The corresponding range of tilt values agrees with the results of
Ref.~\cite{LYTH1}.


\subsection{The potential for $\delta < 1$ }

When $\delta < 1$, instead of a saddle point we have a point of
inflection at $\phi_i$, where $V''(\phi_i) = 0$. We find
\beq \label{infvev}
\phi_i = \phi_0 \left( 1 - {n-2\over32}\alpha^2 +
{\cal O}(\alpha^4) \right)\,,
\eeq
and
\begin{eqnarray}
&&V(\phi_i) =  V(\phi_0)\,\left(1+{3(n-2)(n-4)\over16n}\alpha^2
+ {\cal O}(\alpha^4)\right) \,, \\
&&V'(\phi_i) = \left({n-2\over2}\right)^2
\alpha^2 m^2_{\phi} \phi_0 \, + {\cal O}(\alpha^4) \,,  \label{slope} \\
&&V'''(\phi_i) = V'''(\phi_0)\left( 1 + {\cal O}(\alpha^2)\right) \,.
\end{eqnarray}
The slow-roll parameters at the point of inflection are
\begin{eqnarray} \label{epsinf}
&&\epsilon(\phi_i) = {1\over2}{M_{\rm P}^2\over\phi_0^2}\left({n(n-1)\over2}
\right)^2\,\alpha^4\,,\\
&&\eta(\phi_i) = 0\,.
\end{eqnarray}

Unlike the previous cases, $\delta =1$ (saddle point), and $\delta >
1$ (tunneling solution), there is no point (except the origin
$\phi=0$) where $V^{\prime}=0$. This implies that there will be no
self-reproduction regime unless $\alpha \ll
(m_{\phi}\phi^2_0/M^3_{\rm P})^{1/2}$~\cite{LYTH1}.

However this is not troublesome as long as we have a sufficient number
of e-foldings, ${\cal N} \geq {\cal N}_{\rm COBE}$ arising due to a
slow roll inflation.

The amplitude and tilt of the scalar spectrum in the case $\delta<1$
can be obtained from the analytical continuation $\beta \to i\beta$ of
the results of previous subsection ($\delta>1$),
\begin{eqnarray}
&&\delta_H \simeq {3\over5\pi}\,{H_{\rm inf}^3\over V'(\phi)} =
{(n-2)\sqrt{n}(2n-2)^{n\over2n-4}\over5\pi\sqrt3}\,\lambda_n^{2\over n-2}\,
\left({m_\phi\over M_{\rm P}}\right)^{n-4\over n-2}\,
{\sin^2\beta{\cal N}\over\beta^2}\,,\\[2mm]
&&n_s = 1 - 4\beta\,\cot\beta{\cal N}\,,\\[2mm]
&&{d\,n_s\over d\,\ln k} = - {4\beta^2\over\sin^2\beta{\cal N}}\,,
\end{eqnarray}
which is in agreement with the results of Ref.~\cite{LYTH1}.

The dependence of the tilt on $\beta$ can be seen in
Fig.~\ref{tiltbeta}. Note that, as pointed out in Ref.~\cite{LYTH1},
the tilt can get any value in the allowed range of $\beta$ which is
determined by the viability of slow roll. In the future we will have
to determine what value of $\beta$ agrees with observations.

To summarize the fine tuning issue, for typical values of $\phi_0 \ls
10^{15}$~GeV~\footnote{The tendency from radiative corrections is to
raise $\phi_0$, see Section 4.3.}, the saddle point condition,
Eq.~(\ref{cond}), requires fine-tuning at the level of
\beq \label{fine} \alpha \sim 10^{-9} - 10^{-8}\,, \eeq
which is not negligible.

\section{Radiative and supergravity corrections}

The MSSM inflaton candidates are represented by {\it gauge invariant}
combinations but are not singlets. The inflaton parameters receive
corrections from gauge interactions which, unlike in models with a
gauge singlet inflaton, can be computed in a straightforward way.
Quantum corrections result in a logarithmic running of the soft
supersymmetry breaking parameters $m_\phi$ and $A$. One might then
worry about their impact on Eq.~(\ref{cond}) and the success of
inflation.

In this section we will discuss running of the potential with
VEV-dependent values of $m_\phi(\phi)$ and $A(\phi)$ in
Eq.~(\ref{cond}). Our conclusion is that the running of the gauge
couplings do not spoil the existence of a saddle point. However the
VEV of the saddle point is now displaced; by how much will depend
precisely on the inflaton candidate. In order to discuss the
situation, we derive a general expression for the one-loop effective
potential for the flat directions, and then focus on the ${\bf LLe}$
direction, for which the system of RG equations can be solved
analytically.

\subsection{One-loop effective potential}

The first thing to check is whether the radiative corrections remove
the saddle point altogether. The object of interest is the effective
potential at the phase minimum $n\theta_{\min} = \pi$, for which we
obtain
\begin{eqnarray}
V_{eff}(\phi, \theta_{min}) &=& \frac{1}{2} m_0^2 \phi^2 \left[
  1+ K_1 \log \left(\frac{\phi^2}{\mu_0^2}\right) \right] -
  \frac{\lambda_{n,0} A_0}{nM^{n-3}} \phi^n \left[ 1+ K_2
  \log\left(\frac{\phi^2}{\mu_0^2}\right) \right] \nonumber \\ & & +
  \frac{\lambda_{n,0}^2}{M^{2(n-3)}} \phi^{2(n-1)} \left[ 1 + K_3 \log
  \left(\frac{\phi^2}{\mu_0^2}\right) \right]\,.
\end{eqnarray}
%
%
%
%
%
where $m_0$, $A_0$, and $\lambda_{n,0}$ are the values of $m_{\phi}$,
$A$ and $\lambda_n$ given at a scale $\mu_0$. Here $A_0$ is chosen to
be real and positive (this can always be done by re-parameterizing the
phase of the complex scalar field $\phi$), and $|K_i|<1$ are
coefficients determined by the one-loop renormalization group
equations.

Our aim is to find a saddle point of this effective potential, so we
calculate the 1st and 2nd derivatives of the potential and set them to
zero. This is a straightforward although somewhat cumbersome exercise
that results in the expression
\begin{equation}
\label{phi0-append}
\phi_0^{n-2} = \frac{M^{n-3}}{4\lambda_n(n-1+K_3)} \left[A
\left(
    1+ \frac{2}{n} K_2 \right) \pm \sqrt{A^2 \left( 1+ \frac{2}{n} K_2
    \right)^2 - 8m_{\phi}^2 (1+K_1) (n-1+K_3) } \right]\,,
\end{equation}
where $m_{\phi}$, $A$, and $\lambda_{n}$ are values of the parameters
at the scale $\phi_0$.  Inserting this into $V_{,\phi\phi}=0$, we can
then find the condition to have a saddle point at $\phi_0$:
\begin{eqnarray}
\label{A-append}
&&A^2 = 2 m_{\phi}^2 (n-1+K_3) F_1 F_2 F_3 \, \nonumber \\
&&F_1 = \left[ \frac{1+K_1}{n-1+K_3} \Big( (n-1)(2n-3) +
    (4n-5)K_3 \Big) - 1 - 3K_1 \right]^2 , \, \nonumber \\
&&F_2 = \left[ (1+K_1) \left(n-1+2\frac{2n-1}{n} K_2 \right) - (1+3K_1)
\left( 1+
    \frac{2}{n} K_2 \right) \right]^{-1}, \,  \nonumber \\
&&F_3 =
\left[
\frac{1+\frac{2}{n}K_2}{n-1+K_3} \Big( (n-1)(2n-3) + (4n-5)K_3 \Big)
- \left( n-1+ 2 \frac{2n-1}{n} K_2 \right) \right]^{-1}\,. \nonumber \\
&& \,
\end{eqnarray}
In the limit when $|K_i|\ll 1$, this mercifully simplifies to
\begin{eqnarray} \label{A-append2}
A^2 = 8(n-1) m_{\phi}^2(\phi_0)
\left( 1+ K_1 - \frac{4}{n} K_2 + \frac{1}{n-1} K_3 \right)\,,\\
\phi_0^{n-2} = \frac{M^{n-3} m_{\phi}(\phi_0)}{\lambda_n \sqrt{2(n-1)}}
\left( 1+\frac{1}{2} K_1 - \frac{1}{2(n-1)} K_3 \right)\,.
\end{eqnarray}
Note that Eqs.~(\ref{A-append}),~(\ref{A-append2}) give the necessary
relations between the values of $m_{\phi}$ and $A$ as calculated at
the {\it saddle point}.  The coefficients $K_i$ need to be solved from
the renormalization group equations at the scale given by the saddle
point $\mu=\phi_0$. Since $K_i$ are already one loop corrections,
taking the tree-level saddle point value as the renormalization scale
is sufficient.

Hence we may conclude that, although the soft terms and the value of
the saddle point are all affected by radiative corrections, they do
not remove the saddle point nor shift it to unreasonable values. The
existence of a saddle point is thus insensitive to radiative
corrections.


\subsection{RG equations for the ${\bf L} {\bf L} {\bf e}$ direction}

The form of the relevant RG equations depend on the flat direction. RG
equations for ${\bf LLe}$ are simpler since only the $SU(2)_{W} \times
U(1)_Y$ gauge interactions are involved and the lepton Yukawa
couplings are negligible. The case of ${\bf udd}$ requires numerics if
$u$ is chosen from the third family. For other choices, however, it
closely resembles ${\bf LLe}$. For ${\bf LLe}$ the one-loop RG
equations governing the running of $m^2_{\phi}$, $A$, and $\lambda$
with the scale $\mu$ are given by~\cite{NILLES}
\begin{eqnarray} \label{RGE}
\mu {d m^2_{\phi} \over d \mu} &=& - {1 \over 6 \pi^2} \left({3 \over 2}
{\tilde m_2}^2 g^2_2 + {3 \over 2} {\tilde m_1}^2 g^2_1 \right) \, , \nonumber
\\
\mu {d A \over d \mu} &=& - {1 \over 2 \pi^2} \left({3 \over 2} {\tilde m_2}
g^2_2 + {3 \over 2} {\tilde m_1} g^2_1 \right) \, , \nonumber \\
\mu {d \lambda \over d \mu} &=& - {1 \over 4 \pi^2} \lambda \left({3 \over 2}
g^2_2 + {3 \over 2} g^2_1 \right) \, .
\end{eqnarray}
Here ${\tilde m_1}$, ${\tilde m_2}$ denote the mass of the $U(1)_Y$
and $SU(2)_W$ gauginos respectively and $g_1,~g_2$ are the associated
gauge couplings. It is a straightforward exercise to obtain the
equations that govern the running of $\lambda$ and $A$ associated with
the $\left(LLe\right)^2$ superpotential term (which lifts the ${\bf L}
{\bf L} {\bf e}$ flat direction). Note that ${\bf L}$ has the same
quantum numbers as ${\bf H}_d$, and hence in this respect ${\bf LLe}$
combination behaves just like ${\bf H}_d {\bf L} {\bf e}$. One can
then use the familiar RG equations that govern the Yukawa coupling and
$A$-term associated with the ${H}_d {L} {e}$ superpotential
term~\cite{NILLES}. However, as explained in ~\cite{YAMADA}, the
coefficients of the terms on the right-hand side are proportional to
the number of superfields contained in a superpotential
term.~\footnote{We would like to thank Manuel Drees for explaining this
point to us.} Hence the second and third equations in~(\ref{RGE}) are
simply obtained from those for the ${H}_d {Le}$ term after multiplying
by a factor of $2$. The first equation in~(\ref{RGE}) is also easily
found by taking the electroweak charges of $L_i$, $L_j$ and $e$
superfields into account while taking into account that $m^2_{\phi} =
(m^2_{L_i} + m^2_{L_j}+ m^2_{\bf e})/3$.

The running of gauge couplings and gaugino masses obey the usual
equations~\cite{NILLES}:
\begin{eqnarray}
\mu {d g_1 \over d \mu} &=& {11 \over 16 \pi^2} g^3_1 \, , \nonumber \\
\mu {d g_2 \over d \mu} &=& {1 \over 16 \pi^2} g^3_2 \, , \nonumber \\
{d \over d \mu}\left({{\tilde m_1} \over g_1^2}\right) &=&
{d \over d \mu}\left({{\tilde m_2} \over g_2^2}\right) = 0 \, .
\end{eqnarray}
The solutions of the renormalization group equations are
\begin{eqnarray}
g_i &=& \frac{g_i(\mu_0)}{\sqrt{1-b_i g_i(\mu_0)^2 \ln
    \frac{\mu}{\mu_0} }}, \\
{\tilde m_i} &=& {\tilde m_i}(\mu_0) \left( \frac{g_i}{g_i(\mu_0)}
\right)^2, \\
\label{mphi}
m_{\phi}^2 &=& m_{\phi}^2(\mu_0) + {\tilde m_2}^2(\mu_0) - {\tilde m_2}^2 +
\frac{1}{11} \left( {\tilde m_1}^2(\mu_0) - {\tilde m_1}^2 \right), \\
A &=& A(\mu_0) + 6 \left( {\tilde m_2}(\mu_0) - {\tilde m_2} \right) +
\frac{6}{11} \left( {\tilde m_1}(\mu_0) - {\tilde m_1} \right), \\
\lambda &=& \lambda(\mu_0) \left( \frac{g_2(\mu_0)}{g_2} \right)^6 \left(
  \frac{g_1(\mu_0)}{g_1} \right)^{\frac{6}{11}}~,
\end{eqnarray}
where $i=1,2$,  $b_1=11/8\pi^2$ and $b_2=1/8\pi^2$.  Ignoring the
running of the gaugino masses and gauge couplings, we find that
\begin{eqnarray}
K_1 &\approx& - {1 \over 4 \pi^2} \left[\left({{\tilde m_2}
\over m_{\phi_0}}\right)^2 g^2_2 + \left({{\tilde m_1}
\over m_{\phi_0}}\right)^2 g^2_1 \right] \, , \nonumber \\
K_2 &\approx& - {3 \over 4 \pi^2} \left[\left({{\tilde m_2}
\over A_0}\right) g^2_2 +
\left({{\tilde m_1} \over A_0}\right) g^2_1 \right] \, , \nonumber \\
K_3 &\approx& - {3 \over 8 \pi^2} \lambda_0 \left[g^2_2 + g^2_1 \right] \, ,
\end{eqnarray}
where the subscript $0$ denotes the values of parameters at the high scale
$\mu_0$.

For universal boundary conditions, as in minimal grand unified
supergravity, the high scale is the GUT scale $\mu_X \approx 3 \times
10^{16}$~GeV, ${\tilde m_1}(\mu_X) = {\tilde m_2}(\mu_X) = {\tilde m}$
and $g_1 = \sqrt{\pi/10} \approx 0.56$, $g_2 = \sqrt{\pi/6} \approx
0.72$. Then we just use RG equations to run the coupling constants and masses
to the scale of the saddle point $\mu_0 = \phi_0 \approx 2.6 \times
10^{14}$~GeV for $M_{\rm P} = 2.4 \times 10^{18}$~GeV, $m_{\phi_0}=
1$~TeV, $\lambda_0=1$. With these values we obtain
\begin{eqnarray}
K_1 &\approx & -0.017 \xi^2, \\
K_2 &\approx & -0.0085 \xi, \\
K_3 &\approx & -0.029\,.
\end{eqnarray}
where $\xi = {\tilde m}/ m_{\phi}$ is calculated at the GUT scale.

Typically the running based on gaugino loops alone results in negative
values of $K_i$~\cite{EJM1}. Positive values can be obtained when one
includes the Yukawa couplings, practically the top Yukawa, but the
order of magnitude remains the same.


\begin{figure}
\vspace*{-0.0cm}
\begin{center}
\epsfig{figure=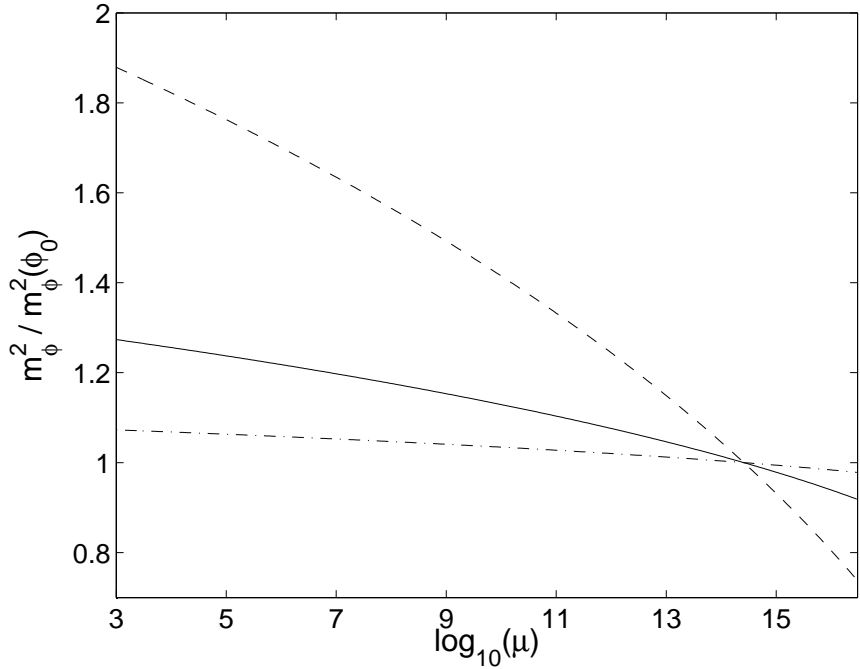,width=.84\textwidth,clip=}
\vspace*{-0.0cm}
\end{center}
\caption{The running of $m^2_{\phi}$ for the $LLe$ inflaton when the
saddle point is at $\phi_0 = 2.6 \times 10^{14}$GeV (corresponding to
$n=6$, $m_{\phi}=1$ TeV and $\lambda = 1$). The three curves
correspond to different values of the ratio of gaugino mass to flat
direction mass at the GUT scale: $\xi = 2$ (dashed), $\xi = 1$ (solid)
and $\xi = 0.5$ (dash-dot).  }\label{LLE-RUN0}
\end{figure}


Thus radiative corrections modify $\alpha$ and we need to finetune the
potential to a few (but not all) orders in perturbation theory.
However, although not completely disastrous, this can hardly be
considered a satisfactory situation, and in the conclusions we
speculate about possible remedies.


\subsection{The inflaton and LHC}

Let us recall that the constraint on the mass of the $n=6$ flat
direction inflaton in Eq.~(\ref{mbound}) reads

\begin{equation}
\label{mbound1} m_{\phi}(\phi_0) \simeq 100\, \rm{GeV} \,
\cdot\lambda^{-1} \, \left( \frac{{\cal N}_{\rm COBE}}{50}
\right)^{-4}\,.
\end{equation}
As mentioned earlier, this is the bound on the mass of the flat
direction during inflation, determined at the scale $\phi=\phi_0$.
Since the inflaton mass runs from $\phi_0$ down to the LHC energy
scales, it will also get scaled.


\begin{figure}
\vspace*{-0.0cm}
\begin{center}
\epsfig{figure=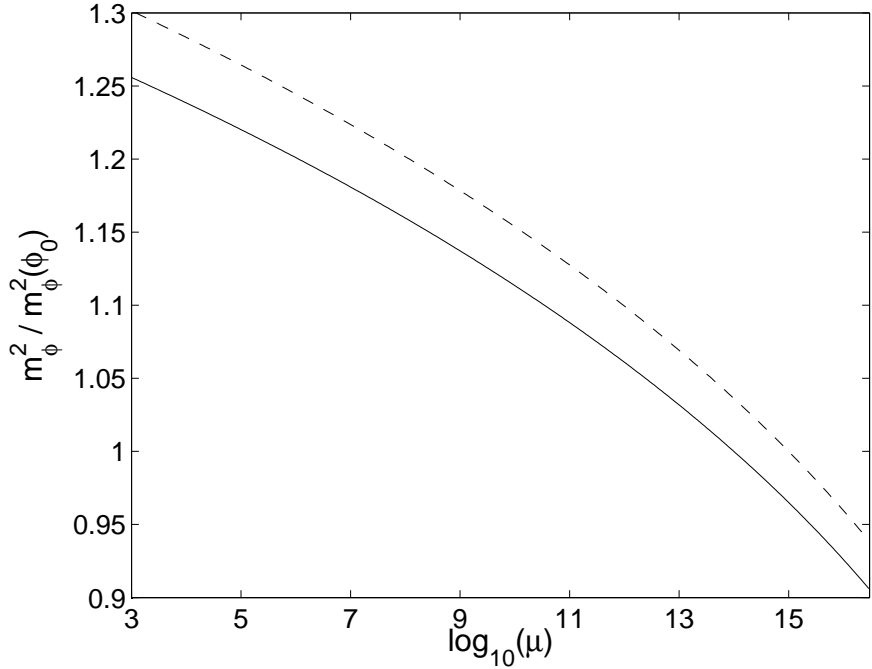,width=.84\textwidth,clip=}
\vspace*{-0.0cm}
\end{center}
\caption{ The same as Fig. 1 but with $\phi_0 =
10^{14}$GeV (solid) and $\phi_0 = 10^{15}$GeV (dashed), and $\xi = 1$.}
\label{LLE-RUN1}
\end{figure}


For ${\bf LLe}$ we see from the solution, Eq. (\ref{mphi}), that the
flat direction mass only gets larger due to the gaugino running,
\beq
\label{mvevin}
m_{\phi}^2 = m_{\phi}^2(\mu_0) + {\tilde m_2}^2(\mu_0) - {\tilde m_2}^2 +
\frac{1}{11} \left( {\tilde m_1}^2(\mu_0) - {\tilde m_1}^2 \right)\,.
\eeq
This has been depicted in Figs.~\ref{LLE-RUN0} and~\ref{LLE-RUN1}.
The $x$-axis is the scale $\mu$ which varies from the saddle point
VEV at $\phi_0$ down to $1$~TeV scale, where LHC is probing new
physics. We have also assumed unification of gauge couplings at the
GUT scale. We find that the mass of the inflaton increases only very
slightly at the TeV scale. In Fig.~\ref{LLE-RUN0} we see that the
increase is $\sqrt{1.9}\times 100,~\sqrt{1.3}\times 100,~\sqrt{1.1}
\times 100$~GeV, for $\xi =2,1,0.5$, respectively. Changing the
initial VEV from $10^{14}$~GeV to $10^{15}$~GeV results only in a
minor modification in the running of $m^2_{\phi}$.  This has been
depicted in Fig.~\ref{LLE-RUN1}.

The mass at the TeV scale will increase further if $\lambda$
decreases much below one. For $\lambda=10^{-2}$, the bound exceeds
$10$~TeV which escapes LHC limit. On the other hand, note that LHC
will never probe the flat direction mass directly, but may set a
limit on the slepton masses. However, we do not claim that LHC can
discover MSSM inflation, but it can certainly rule out the
possibility. If LHC does not find low energy supersymmetry within
$\sim $~TeV, then MSSM inflation is effectively ruled out.

The situation would be similar for ${\bf udd}$ without the top squark.
For the ${\bf u_3dd}$ direction it is possible that the inflaton mass
gets even smaller at the weak scale.

\subsection{ $A_6$~vs.~$A_3$}

One final comment is in order before closing this Section. Unlike
$m_{\phi}$, there is no prospect of measuring the $A$ term, because it
is related to the non-renormalizable interactions which are suppressed
by $M_{\rm P}$. However, a knowledge of supersymmetry breaking sector
and its communication with the observable sector may help to link the
non-renormalizable $A$-term under consideration to the renormalizable
ones.

To elucidate this, let us consider the Polonyi model where a general
$A$-term at a tree level is given by
$$m_{3/2}[(a-3)W+\phi (dW/d\phi)],$$
with $a=3 - \sqrt{3}$~\cite{NILLES}. One then finds a relationship
between $A$-terms corresponding to $n=6$ and $n=3$ superpotential
terms, denoted by $A_6$ and $A_3$ respectively, at high scales:
\beq \label{polon}
A_6={3 - \sqrt{3} \over 6 - \sqrt{3}} A_3\,.
\eeq
One can then use relevant RG equations to relate $A_6$ which is relevant for
inflation, to $A_3$ at the weak scale, which can be constrained and/or
measured. In principle this can also be done in general, provided that
we have sufficient information about the supersymmetry breaking sector
and its communication with the MSSM sector.


\subsection{Supergravity corrections}

SUGRA corrections often destroy the slow roll predictions of
inflationary potentials; this is the notorious SUGRA-$\eta$
problem~\cite{ETA}. In general, the effective potential depends on the
K\"ahler potential $K$ as $ V\sim
\left(e^{K(\varphi^{\ast},\varphi)/M_{\rm P}^2} V(\phi)\right) $ so
that there is a generic SUGRA contribution to the flat direction
potential of the type
\begin{equation}
\label{mflat}
V(\phi)=H^2M_{\rm P}^2 f\left(\frac{\phi}{M_{\rm P}}\right)\,,
\end{equation}
where $f$ is some function (typically a polynomial).  Such a
contribution usually gives rise to a Hubble induced correction to the
mass of the flat direction with an unknown coefficient, which depends
on the nature of the K\"ahler potential~\footnote{If the K\"ahler
potential has a shift symmetry, then at tree level there is no Hubble
induced correction. However, at one-loop level relatively small Hubble
induced corrections can be induced~\cite{GMO,ADM}.}.

Let us compare the non-gravitational contribution, Eq.~(\ref{scpot}),
to that of Hubble induced contribution, Eq.~(\ref{mflat}). Writing
$f\sim \left( \phi/M_{\rm P}\right)^p$ where $p\ge 1$ is some power,
we see that non-gravitational part dominates whenever
\beq
H_{\rm inf}^2M_{\rm P}^2\left(\frac{\phi}{M_{\rm P}}\right)^p \ll
m_{\phi}^2\phi_0^2\,,
\eeq
so that the SUGRA corrections are negligible as long as $\phi_0 \ll
M_{\rm P}$, as is the case here (note that $H_{\rm inf} M_{\rm P} \sim
m_{\phi} \phi_0$).  The absence of SUGRA corrections is a generic
property of this model. Note also that although non-trivial K\"ahler
potentials give rise to non-canonical kinetic terms of squarks and
sleptons, it is a trivial exercise to show that at sufficiently low
scales, $H_{\rm inf}\ll m_{\phi}$, and small VEVs, they can be rotated
to a canonical form without affecting the potential~\footnote{The same
reason, i.e. $H_{\rm inf}\ll m_{\phi}$ also precludes any large
Trans-Planckian correction. Any such correction would generically go
as $(H_{\rm inf}/M_{\ast})^2\ll 1$, where $M_{\ast}$ is the scale at which
one would expect Trans-Planckian effects to kick in~\cite{BCH}.}.


\section{End of MSSM inflation}
\subsection{Reheating and Thermalization}

After the end of inflation, the flat direction starts rolling towards
its global minimum. At this stage the dominant term in the scalar
potential will be: $m_\phi \phi^2/2$. Since the frequency of
oscillations is $\omega \sim m_{\phi} \sim 10^3 H_{\rm inf}$, the flat
direction oscillates a large number of times within the first Hubble
time after the end of inflation. Hence the effect of expansion is
negligible.

We recall that the curvature of the potential along the angular
direction is much larger than $H^2_{\rm inf}$. Therefore, the flat
direction has settled at one of the minima along the angular direction
during inflation from which it cannot be displaced by quantum
fluctuations. This implies that no torque will be exerted, and hence
the flat direction motion will be one dimensional, i.e. along the
radial direction.

Flat direction oscillations excite those MSSM degrees of freedom which
are coupled to it.  The inflaton, either ${\bf LLe}$ or ${\bf u}{\bf
d}{\bf d}$ flat direction, is a linear combination of slepton or
squark fields. Therefore inflaton has gauge couplings to the
gauge/gaugino fields and Yukawa couplings to the Higgs/Higgsino
fields. As we will see particles with a larger couplings are produced
more copiously during inflaton oscillations. Therefore we focus on the
production of gauge fields and gauginos. Keep in mind that the VEV of
the MSSM flat direction breaks the gauge symmetry spontaneously, for
instance ${\bf udd}$ breaks $SU(3)_C \times U(1)_Y$ while ${\bf LLe}$
breaks $SU(2)_{W}\times U(1)_{Y}$, therefore, induces a supersymmetry
conserving mass $\sim g \langle \phi(t) \rangle$ to the gauge/gaugino
fields in a similar way as the Higgs mechanism, where $g$ is a gauge
coupling. When the flat direction goes to its minimum, $\langle
\phi(t)\rangle = 0$, the gauge symmetry is restored. In this respect
the origin is a point of enhanced symmetry~\cite{AVERDI2}.

There can be various phases of particle creation in this model, here
we briefly summarize them below.  Let us elucidate the physics, by
considering the case when ${\bf LLe}$ flat direction is the
inflaton.

\begin{itemize}

\item{Tachyonic preheating:\\ Right after the end of inflation, when
we are close to the saddle point, the second derivative is negative.
One might suspect that this would trigger tachyonic instability in the
inflaton fluctuations which will then excite the inflaton couplings to
matter~\cite{NEWPREH,ESTER}.

However the situation is different in our case. As mentioned, only
inflaton fluctuations with a physical momentum $k \ls m_{\phi}$ will
have a tachyonic instability.  Moreover $V^{\prime \prime} < 0$ only
at field values which are $\sim \phi_0$.  Tachyonic effects are
therefore expected be negligible since, unlike the case
in~\cite{NEWPREH}, the homogeneous mode has a VEV which is
hierarchically larger than $m_{\phi}$ (we remind that $\phi_0 \geq
10^{14}$ GeV) and oscillates at a frequency $\omega \sim
m_{\phi}$. Further note fields which are coupled to the inflaton
acquire a very large mass $\sim h \phi_0$ from the homogeneous piece
which suppresses non-perturbative production of their quanta at large
inflaton VEVs. We conclude that tachyonic effects, although genuinely
present, do not lead to significant particle production in our case.
}

\item{Instant preheating:\\ An efficient bout of particle creation
occurs when the inflaton crosses the origin, which happens twice in
every oscillation. The reason is that fields which are coupled to the
inflaton are massless near the point of enhanced symmetry. Mainly
electroweak gauge fields and gauginos are then created as they have
the largest coupling to the flat direction. The production takes place
in a short interval, $\Delta t \sim \left(g m_{\phi} \phi_0
\right)^{-1/2}$, where $\phi_0\sim 10^{14}$~GeV is the initial
amplitude of the inflaton oscillation, during which quanta with a
physical momentum $k \ls \left(g m_{\phi} \phi_0 \right)^{1/2}$ are
produced. The number density of gauge/gaugino degrees of freedom is given
by~\cite{PREHEAT}
\beq \label{chiden}
n_{g} \approx {\left(g m_{\phi} \phi_0
\right)^{3/2} \over 8 \pi^3}\,.
\eeq
As the inflaton VEV is rolling back to its maximum value $\phi_0$, the
mass of the produced quanta $g \langle \phi(t) \rangle$ increases. The
gauge and gaugino fields can (perturbatively) decay to the fields
which are not coupled to the inflaton, for instance to (s)quarks. Note
that (s)quarks are not coupled to the flat direction, hence they
remain massless throughout the oscillations. The total decay rate of
the gauge/gaugino fields is then given by $\Gamma = C \left(g^2/48\pi
\right) g\phi $, where $C \sim {\cal O}(10)$ is a numerical factor
counting for the multiplicity of final states.

The decay of the gauge/gauginos become efficient when
\beq \label{ddecay}
\langle \phi \rangle \simeq \left({48 \pi m_{\phi} \phi_0 \over
C g^3}\right)^{1/2}\,.
\eeq
Here we have used $\langle \phi(t) \rangle \approx \phi_0 m_{\phi} t$,
which is valid when $m_{\phi} t \ll 1$, and $\Gamma \simeq t^{-1}$,
where $t$ represents the time that has elapsed from the moment that
the inflaton crossed the origin. Note that the decay is very quick
compared with the frequency of inflaton oscillations, i.e. $\Gamma \gg
m_{\phi}$. It produces relativistic (s)quarks with an energy:
\beq \label{energy}
E =\frac{1}{2}g\phi(t)
\simeq \left({48 \pi m_{\phi} \phi_0 \over C g}\right)^{1/2}\,.
\eeq
The ratio of energy density in relativistic particles thus produced
$\rho_{rel}$with respect to the total energy density $\rho_0$ follows
from Eqs.~(\ref{chiden}),~(\ref{energy}):
\beq
\label{ratio}
{\rho_{rel} \over \rho_0} \sim 10^{-2} g\,,
\eeq
where we have used $C \sim {\cal O}(10)$.  This implies that a
fraction $\sim {\cal O}(10^{-2})$ of the inflaton energy density is
transferred into relativistic (s)quarks every time that the inflaton
passes through the origin. This is so-called instant preheating
mechanism~\cite{INSTANT}~\footnote{In a favorable condition the flat
direction VEV coupled very weakly to the flat direction inflaton
could also enhance the perturbative decay rate of the
inflaton~\cite{ABM}.}. It is quite an efficient mechanism in our
model as it can convert almost all of the energy density in the
inflaton into radiation within a Hubble time (note that $H^{-1}_{\rm
inf} \sim 10^3 m^{-1}_{\phi}$)~\footnote{We emphasize that reheating
happens quickly due to a flat direction motion which is {\it
strictly} one dimensional in our case. Our case is really
exceptional; usually, the flat direction motion is restricted to a
plane, which precludes preheating all together, for instance
see~\cite{Postma}.}.}

\end{itemize}

\subsection{Towards thermal equilibrium}

A full thermal equilibrium is reached when ${\it a)~kinetic }$ and
${\it b)~chemical~equilibrium }$ are established. The maximum
(hypothetical) temperature attained by the plasma would be given by:
\beq
\label{tmax}
T_{max} \sim V^{1/4} \sim \left(m_{\phi}\phi_0\right)^{1/2}
\geq 10^{9}~{\rm GeV}\,.
\eeq
This temperature may be too high and could lead to thermal
overproduction of gravitinos~\cite{Ellis,Buchmuller}. However the
dominant source of gravitino production in a thermal bath is
scattering which include an on-shell gluon or gluino leg. In the next
subsection we describe a natural solution to this problem and show that
 the final reheat temperature is actually well below
Eq.~(\ref{tmax}), i.e. $T_{R}\ll T_{max}$.

One comment is in order before closing this subsection. The gravitinos
can also be created non-perturbatively during inflaton oscillations,
both of the helicity $\pm 3/2$~\cite{MAROTO} and helicity $\pm 1/2$
states~\cite{REST}. In models of high scale inflation (i.e. $H_{\rm
inf} \gg m_{3/2}$) helicity $\pm 1/2$ states can be produced very
efficiently (and much more copiously than helicity $\pm 3/2$ states).
At the time of production these states mainly consist of the inflatino
(inflaton's superpartner).  However these fermions also decay in the
form of inflatino, which is coupled to matter with a strength which is equal
to that of the inflaton. Therefore, they inevitably decay at a similar
rate as that of inflaton, and hence pose no threat to primordial
nucleosynthesis~\cite{MAR}.

In the present case $m_{\phi} \sim m_{3/2} \gg H_{\rm inf}$. Therefore
low energy supersymmetry breaking is dominant during inflation,
and hence helicity $\pm 1/2$ states of the gravitino are not related
to the inflatino (which is a linear combination of leptons or
quarks)at any moment of time. As a result helicity $\pm 1/2$ and $\pm
3/2$ states are excited equally, and their abundances are suppressed
due to kinematical phase factor.  Moreover there will be no dangerous
gravitino production from perturbative decay of the inflaton
quanta~\cite{AEM,SURFACE}. The reason is that the inflaton is not a
{\it gauge singlet} and has gauge strength couplings to other MSSM
fields. This makes the $inflaton \rightarrow inflatino ~+~ gravitino$
~decay mode totally irrelevant.


\subsection{Solution to the gravitino problem}

In order to suppress thermal gravitino production it is
sufficient to make gluon and gluino fields heavy enough such that they are
not kinematically accessible to the reheated plasma, even if other
degrees of freedom reach full equilibrium (for a detailed discussion
on thermalization in supersymmetric models and its implications,
see~\cite{AVERDI1,AVERDI2}). This suggests a natural solution to the thermal gravitino problem
in the
case of our model. Consider another flat direction with a
non-zero VEV, denoted by $\varphi$, which spontaneously breaks the
$SU(3)_C$ group.  For example, if ${\bf LLe}$ is the inflaton, then
${\bf udd}$ provides a unique candidate which can simultaneously
develop VEV~\footnote{To develop and maintain such a large VEV, it is
not necessary that ${\bf udd}$ potential has a saddle point as
well. It can be trapped in a false minimum during inflation, which
will then be lifted by thermal corrections when the inflaton decays
(as discussed in the previous subsection)~\cite{AEJM}.}The induced
mass for gluon/gluino fields will be:
\beq \label{flatvev}
m_{G} \sim  g \langle \varphi(t) \rangle < g \phi_0\,.
\eeq
The inequality arises due to the fact that the VEV of $\varphi$ cannot
exceed that of the inflaton $\phi$ since its energy density should be
subdominant to the inflaton energy density.

If $g \varphi_0 \gg T_{max}$ the gluon/gluino fields will be too heavy
and not kinematically accessible to the reheated plasma. Here
$\varphi_0$ is the VEV of ${\bf udd}$ at the beginning of inflaton
oscillations.  In a radiation-dominated Universe the Hubble expansion
redshifts the flat direction VEV as $\langle \varphi \rangle \propto
H^{3/4}$, which is a faster rate than the change in the temperature $T
\propto H^{1/2}$. Once $g \langle \varphi \rangle \simeq T$,
gluon/gluino fields come into equilibrium with the thermal bath.  As
pointed out in Refs.~\cite{AVERDI1,AVERDI2}, if the initial VEV of
${\bf udd}$ is
\beq
\label{reqVEV}
\varphi_0 > 10^{10}~{\rm GeV}\,,
\eeq
then the temperature at which gluon/gluino become kinematically
accessible, i.e. $g \langle \varphi \rangle \simeq T$, is given by
~\cite{AVERDI2}~\footnote{Note that the conditions in
Eqs.~(\ref{flatvev}),~(\ref{reqVEV}) can be simultaneously satisfied
easily.}:
\beq
T_{\rm R} \leq 10^{7}~{\rm GeV}\,.
\eeq
This is the final reheat temperature at which gluons and
gluinos are all in thermal equilibrium with the other degrees
of freedom. The standard calculation of thermal gravitino production via
scatterings can then be used for $T \leq T_{\rm R}$. Note however that $T_{\rm R}$
is sufficiently low to avoid thermal overproduction of gravitinos.

Finally, we also make a comment on the cosmological moduli problem.
The moduli are generically displaced from their true
minimum if their mass is less than the expansion rate during
inflation. The moduli obtain a mass $\sim {\cal O}({\rm TeV})$ from
supersymmetry breaking. They start oscillating with a large amplitude,
possibly as big as $M_{\rm P}$, when the Hubble parameter drops below
their mass. Since moduli are only gravitationally coupled to other
fields, their oscillations dominate the Universe while they decay very
late.  The resulting reheat temperature is below MeV, and is too low
to yield a successful primordial nucleosynthesis.

However, in our case $H_{\rm inf} \ll {\rm TeV}$ . This
implies that quantum fluctuations cannot displace the moduli from
their true minima during the inflationary epoch driven by MSSM flat
directions. Moreover, any oscillations of the moduli will be
exponentially damped during the inflationary epoch. Therefore our
model is free from the infamous moduli problem.


\section{Conclusion}

The existence of a saddle point in the scalar potential of the ${\bf
udd}$ or ${\bf LLe}$ MSSM flat directions appears, perhaps
surprisingly, to provide all the necessary ingredients for an
observationally realistic model of inflation~\cite{AEGM}. MSSM
inflation takes place at a low energy scale so that it is naturally
free of supergravity and super-Planckian effects. The exceptional
feature of the model, which sets it apart from conventional singlet
field inflation models, is the fact that here the inflaton is a gauge
invariant combination of the squark or slepton fields. As a
consequence, the couplings of the inflaton to the MSSM matter and
gauge fields are known.  This makes it possible to address the
questions of reheating and gravitino production in an unambiguous way,
as we did in Sect. 5.  Since ${\bf udd}$ and ${\bf LLe}$ are
independently flat, therefore, if ${\bf LLe}$ is the inflaton, the
${\bf udd}$ direction can also acquire a large VEV simultaneously.
This gives a large mass to gluons/gluinos which decouples them from
the thermal bath, and hence suppresses thermal gravitino
production. As discussed in Sect. 5, non-thermal production of
gravitinos is negligible in our model.

In the MSSM inflation model the mass of the inflaton is not a free
parameter but is related to the masses of e.g. sleptons, should the
${\bf LLe}$ direction be the inflaton. We have solved the appropriate
RG equation equations to relate the inflaton mass to the slepton
masses at energies accessible to accelerators such as LHC and found
that LHC can indeed put a constraint on the model: it may not be able
to verify it, but it certainly can rule it out.

The model predictions are not modified by supergravity corrections,
i.e. the observables are insensitive to the nature of K\"ahler
potential. MSSM inflation also illustrates that it is free from any
Trans-Planckian corrections.  MSSM inflation retains the successes of
thermal production of LSP as a dark matter and the electroweak
baryogenesis within MSSM.

The existence of the saddle point requires a fine-tuning of ratio of
the soft breaking terms $A$ and $m$, or the parameter $\alpha$, as
discussed in length in Sects. 3 and 4. We dealt both with the case of
a local minimum and the case of $V'(\phi)>0$.  We found that a
fine-tuning of the order of $10^{-9}$ is sufficient.  It is therefore
necessary to adjust the ratio $A/m$ up to few orders in perturbation
theory.  However, we find that the existence of the saddle point is
not sensitive to radiative corrections so that saddle point inflation
can always be achieved for some value of the ratio $A/m$.

However, it is conceivable that the mechanism of supersymmetry
breaking, which lies outside the effective theory of MSSM combined
with gravity discussed in this paper, could remove the fine-tuning
in some natural, dynamical way. For instance, $A/m$ could turn out
to be a renormalization group fixed point so that once the ratio is
fixed, it would remain fixed at all orders (for example,
see~\cite{ROSS}). This requires a detailed investigation, but it is
warranted by the simplicity and the apparent success of MSSM flat
direction inflation, which is unique in being both a successful
model of inflation and at the same time having a concrete and real
connection to physics that can be observed in earth bound
laboratories.


\section{Acknowledgments}

We wish to thank Cliff Burgess, Manuel Drees, John Ellis, Jaume
Garriga, Alex Kusenko, and Tony Riotto for valuable discussions and
various suggestions they have made. We also benefitted from the
discussions with Shanta de Alwis, Steve Abel, Mar Bastero-Gil, Micha
Berkhooz, Zurab Berezhiani, Robert Brandenberger, Ramy Brustein,
Damien Easson, Gordy Kane, Justin Khoury, George Lazarides, Andrei
Linde, Andrew Liddle, Hans Peter Nilles, Pavel Naselsky, Maxim
Pospelov, Subir Sarkar, Qaisar Shafi, Misha Shaposhnikov, Scott
Thomas and Igor Tkachev. We would also like to thank the Galileo
Galilei Institute for Theoretical Physics for the hospitality and
the INFN for partial support during the completion of this work. The
research of RA was supported by Perimeter Institute for Theoretical
Physics. Research at Perimeter Institute is supported in part by the
Government of Canada through NSERC and by the province of Ontario
through MEDT. KE is supported by the Academy of Finland grant
108712. The research of KE, AJ, JGB and AM are partly supported by
the European Union through Marie Curie Research and Training Network
``UNIVERSENET'' (MRTN-CT-2006-035863).



\begin{thebibliography}{50}


\bibitem{AEGM}
R.~Allahverdi, K.~Enqvist, J.~Garcia-Bellido and A.~Mazumdar,
  ``Gauge invariant MSSM inflaton,''
  Phys.\ Rev.\ Lett.\ {\bf 97}, 191304 (2006)
  [hep-ph/0605035].


\bibitem{KARI-REV}
K.~Enqvist and A.~Mazumdar,
  Phys.\ Rept.\  {\bf 380}, 99 (2003)
  [hep-ph/0209244].
M.~Dine and A.~Kusenko,
  Rev.\ Mod.\ Phys.\  {\bf 76}, 1 (2004)
  [hep-ph/0303065].



\bibitem{AKM}
 R.~Allahverdi, A.~Kusenko and A.~Mazumdar,
  arXiv:hep-ph/0608138.

\bibitem{LYTH}
 D.~H.~Lyth and A.~Riotto,
  Phys.\ Rept.\  {\bf 314}, 1 (1999)
  [hep-ph/9807278].


\bibitem{AVERDI1}
R.~Allahverdi and A.~Mazumdar,
  ``Towards a successful reheating within supersymmetry,''
  arXiv:hep-ph/0603244.


\bibitem{AVERDI2}
R.~Allahverdi and A.~Mazumdar,
  JCAP\ {\bf 0610}, 008 (2006)
  [hep-ph/0512227].


\bibitem{AVERDI3}
R.~Allahverdi and A.~Mazumdar,
  ``Quasi-thermal universe and its implications for gravitino production,
baryogenesis and dark matter,''
  arXiv:hep-ph/0505050.


\bibitem{AVERDI4}
 R.~Allahverdi and A.~Mazumdar,
  ``Longevity of supersymmetric flat directions,''
  arXiv:hep-ph/0608296.


\bibitem{LYTH1}
 J.~C.~B.~Sanchez, K.~Dimopoulos and D.~H.~Lyth,
  JCAP\ {\bf 0701}, 015 (2007)
  [hep-ph/0608299].


\bibitem{LYTH0}
D. H. Lyth, arXiv:hep-ph/0605283.

\bibitem{GKM}
 T.~Gherghetta, C.~F.~Kolda and S.~P.~Martin,
  Nucl.\ Phys.\ B {\bf 468}, 37 (1996)
  [hep-ph/9510370].


\bibitem{DRT}
M.~Dine, L.~Randall and S.~Thomas, Phys.\ Rev.\ Lett.\ {\bf 75}, 398 (1995)
  [hep-ph/9503303];
  Nucl.\ Phys.\ B\ {\bf 458}, 291 (1996)
  [hep-ph/9507453].



\bibitem{AEJM}
R.~Allahverdi, K.~Enqvist, A.~Jokinen and A.~Mazumdar,
  ``Identifying the curvaton within MSSM,''
  JCAP\ {\bf 0610}, 007 (2006)
  [hep-ph/0603255].


\bibitem{GUTH}
A.~H.~Guth,
  Phys.\ Rev.\ D {\bf 23}, 347 (1981).

\bibitem{JOKINEN}
A.~Jokinen and A.~Mazumdar,
  Phys.\ Lett.\ B {\bf 597}, 222 (2004)
  [hep-th/0406074].



\bibitem{GMSB}
G.~F.~Giudice and R.~Rattazzi,
  Phys.\ Rept.\  {\bf 322}, 419 (1999)
  [hep-ph/9801271].



\bibitem{SPLIT}
N.~Arkani-Hamed, S.~Dimopoulos, G.~F.~Giudice and A.~Romanino,
  Nucl.\ Phys.\ B {\bf 709}, 3 (2005)
  [hep-ph/0409232].


\bibitem{AJKM}
R.~Allahverdi,~A. Jokinen, and A.~Mazumdar,
  ``Sub-eV Hubble scale inflation within GMSB'',
  [hep-ph/0610243].

\bibitem{EJM}
K.~Enqvist, A.~Jokinen and A.~Mazumdar,
  JCAP {\bf 0401}, 008 (2004)
  [hep-ph/0311336].


\bibitem{MULTI}
C.~P.~Burgess, R.~Easther, A.~Mazumdar, D.~F.~Mota and T.~Multamaki,
  JHEP {\bf 0505}, 067 (2005)
  [hep-th/0501125].



\bibitem{Maz-Roz}
R.~Allahverdi and A. Mazumdar,
``Spectral tilt in A-term inflation,'' arXiv:hep-ph/0610069.

\bibitem{WMAP3}
D.N. Spergel, et.al., astro-ph/0603449.


\bibitem{COLEMAN}
R.~Coleman and F.~De Luccia,
  Phys.\ Rev.\ D {\bf 21}, 3305 (1980).


\bibitem{NILLES}
 H.~P.~Nilles,
  Phys.\ Rept.\  {\bf 110}, 1 (1984).


\bibitem{YAMADA}
Y. Yamada, Phys. Rev. D {\bf 50}, 3537 (1995) [hep-ph/9401241].


\bibitem{EJM1}
K.~Enqvist, A.~Jokinen and J.~McDonald,
  Phys.\ Lett.\ B {\bf 483}, 191 (2000)
  [hep-ph/0004050].


\bibitem{ETA}
M. Dine, W. Fischler, and D. Nemeschansky,
   Phys. Lett. B {\bf 136}, 169 (1984);
G. D. Coughlan, R. Holman, P. Ramond, and G. G. Ross,
   Phys. Lett. B {\bf 140}, 44 (1984);
A. S. Goncharov, A. D. Linde, and M. I. Vysotsky,
   Phys. Lett. B {\bf 147}, 279 (1984);
O. Bertolami, and G. G. Ross,
   Phys. Lett. B {\bf 183}, 163 (1987);
E. J. Copeland, A. R. Liddle, D. H. Lyth, E. D. Stewart, and D. Wands,
   Phys. Rev. D {\bf 49}, 6410 (1994).


\bibitem{GMO}
M.~K.~Gaillard, H.~Murayama and K.~A.~Olive,
   Phys. Lett. B {\bf 355}, 71 (1995) [hep-ph/9504307].



\bibitem{ADM}
R. Allahverdi, M. Drees and A. Mazumdar,
   Phys. Rev. D {\bf 65}, 065010 (2002) [hep-ph/0108225].

\bibitem{BCH}
See for instance, C.~P.~Burgess, J.~Cline and R.~Holman,
   JCAP {\bf 0310}, 004 (2003) [hep-th/0306079].


\bibitem{NEWPREH}
G.~N.~Felder, J.~Garcia-Bellido, P.~B.~Greene, L.~Kofman, A.~D.~Linde and
I.~Tkachev,
  Phys.\ Rev.\ Lett.\  {\bf 87}, 011601 (2001)
  [hep-ph/0012142].


\bibitem{Postma}
R.~Allahverdi, R.~H.~A.~Shaw and B.~A.~Campbell,
  Phys.\ Lett.\ B {\bf 473}, 246 (2000)
 [hep-ph/9909256];
 M.~Postma and A.~Mazumdar,
  JCAP {\bf 0401}, 005 (2004)
 [hep-ph/0304246].

\bibitem{ESTER}
J.~Garcia-Bellido and E.~Ruiz Morales,
  Phys.\ Lett.\ B {\bf 536}, 193 (2002)
  [hep-ph/0109230].


\bibitem{PREHEAT}
L.~Kofman, A.~D.~Linde and A.~A.~Starobinsky,
  Phys.\ Rev.\ Lett.\  {\bf 73}, 3195 (1994)
  [hep-th/9405187];
L.~Kofman, A.~D.~Linde and A.~A.~Starobinsky,
  Phys.\ Rev.\ D {\bf 56}, 3258 (1997)
  [hep-ph/9704452].


\bibitem{INSTANT}
G.~N.~Felder, L.~Kofman and A.~D.~Linde,
  Phys.\ Rev.\ D {\bf 59}, 123523 (1999)
  [hep-ph/9812289].




\bibitem{ABM}
R.~Allahverdi, R.~Brandenberger and A.~Mazumdar,
  Phys.\ Rev.\ D {\bf 70}, 083535 (2004)
  [hep-ph/0407230].



\bibitem{Ellis}
J.~R.~Ellis, J.~E.~Kim and D.~V.~Nanopoulos,
  Phys.\ Lett.\ B {\bf 145}, 181 (1984).

\bibitem{Buchmuller}
M.~Bolz, A.~Brandenburg and W.~Buchm\"uller,
Nucl.\ Phys.\ B {\bf 606}, 518 (2001)
[hep-ph/0012052].

\bibitem{MAROTO}
A.~L.~Maroto and A.~Mazumdar,
Phys.\ Rev.\ Lett.\  {\bf 84}, 1655 (2000)
[hep-ph/9904206].

\bibitem{REST}
R.~Kallosh, L.~Kofman, A.~D.~Linde and A.~Van Proeyen,
Phys.\ Rev.\ D {\bf 61}, 103503 (2000)
[hep-th/9907124].


\bibitem{MAR}
R.~Allahverdi, M.~Bastero-Gil and A.~Mazumdar,
Phys.\ Rev.\ D {\bf 64}, 023516 (2001)
[hep-ph/0012057].
H.~P.~Nilles, M.~Peloso and L.~Sorbo,
Phys.\ Rev.\ Lett.\  {\bf 87}, 051302 (2001)
[hep-ph/0102264].
H. P. Nilles, M. Peloso and L. Sorbo,
JHEP {\bf 0104}, 004 (2001) [hep-th/0103202].


\bibitem{AEM}
R.~Allahverdi, K.~Enqvist and A.~Mazumdar,
  Phys.\ Rev.\ D {\bf 65}, 103519 (2002)
  [hep-ph/0111299].


\bibitem{SURFACE}
K.~Enqvist, S.~Kasuya and A.~Mazumdar,
  Phys.\ Rev.\ Lett.\  {\bf 89}, 091301 (2002)
  [hep-ph/0204270].
K.~Enqvist, S.~Kasuya and A.~Mazumdar,
  Phys.\ Rev.\ D {\bf 66}, 043505 (2002)
  [hep-ph/0206272].


\bibitem{ROSS}

M. Lanzagorta and G. G. Ross, Phys. Lett. B {\bf 349}, 319 (1995)
[hep-ph/9501394].
M. Lanzagorta and G. G. Ross, Phys. Lett. B {\bf 364}, 163 (1995)
[hep-ph/9507366].






\end{thebibliography}
\end{document}